\documentclass[3p]{elsarticle}

\usepackage{hyperref}
\usepackage{amsmath}
\usepackage{amsfonts}
\usepackage{amssymb}
\usepackage{graphicx}

\newcommand{\D}{{\rm d}}

\usepackage{xcolor}
\usepackage{soul}

\biboptions{sort&compress}
\begin{document}

\begin{frontmatter}

\title{Modeling and Simulation of Financial Returns \\ under Non-Gaussian Distributions}

\author[mymainaddress,mysecondaryaddress]{Federica De Domenico}
\author[mymainaddressg]{Giacomo Livan}
\author[mymainaddress,mysecondaryaddress]{Guido Montagna\corref{mycorrespondingauthor}}
\cortext[mycorrespondingauthor]{Corresponding author}
\ead{guido.montagna@unipv.it}
\author[mysecondaryaddress]{Oreste~Nicrosini}
\address[mymainaddress]{Dipartimento di Fisica, Universit\`a degli Studi di Pavia, Via A. Bassi 6, 27100, Pavia, Italy}
\address[mysecondaryaddress]{Istituto Nazionale di Fisica Nucleare, Sezione di Pavia, Via A. Bassi 6, 27100, Pavia, Italy}
\address[mymainaddressg]{Department of Computer Science, University College London, 66-72 Gower Street, London~WC1E 6EA, United Kingdom}

\begin{abstract}
It is well known that the probability distribution of high-frequency financial returns is
characterized by a leptokurtic, heavy-tailed shape. This behavior 
undermines the typical assumption of Gaussian log-returns behind the standard 
approach to risk management and option pricing. Yet, there is no consensus 
on what class of probability distributions should be adopted to describe financial 
returns and different models used in the literature have demonstrated, to varying extent, 
an ability to reproduce empirically observed stylized facts. In order to provide 
some clarity, in this paper we perform a thorough study 
of the most popular models of return distributions as obtained in the empirical analyses of 
high-frequency financial data. We compare the 
statistical properties and simulate the dynamics of non-Gaussian 
financial fluctuations by 
means of Monte Carlo sampling from the different models in terms of realistic tail exponents. 
Our findings show a noticeable consistency between the considered 
return distributions in the modeling of the scaling properties 
of large price changes. We also discuss the convergence rate to 
the asymptotic distributions of the non-Gaussian
stochastic processes and we study, as a 
first example of possible applications, 
the impact of our results on option pricing in comparison
with the standard Black and Scholes approach.
\end{abstract}


\end{frontmatter}

\section{Introduction}

Since the seminal papers of Mandelbrot~\cite{Mandelbrot1963} and Fama~\cite{Fama1965} on the statistical 
properties of commodity and stock market prices,
many studies in finance and econophysics over the last 
few decades clearly showed that the empirical distribution of financial 
log-returns deviates from the Gaussian shape inherent to the Brownian motion dynamics
introduced by Osborne~\cite{Osborne1959} and Samuelson~\cite{Samuelson1965}, 
 following Bachelier~\cite{Bachelier1900}. The deviation 
of the empirical distributions from the 
bell-shaped curve is particularly pronounced in the high-frequency limit of 
intraday returns~\footnote{In the paper, we use returns and log-returns as 
synonyms when referring to financial data and associated modeling, since it is known 
that they behave as equivalent empirical random variables in the high-frequency regime~\cite{MSBook2000,BPBook2003}. 
Distinction is made in the simulation of their
stochastic dynamics and study of the limit theorems, as detailed in the following.} and is a universal 
feature of the financial market dynamics, being observed 
across different markets, speculative prices and epochs~\cite{MSBook2000,BPBook2003,Cont2001,Jondeau2006,Gabaix2009}.

The empirical return distribution typically exhibits a sharp central body and a fat-tailed 
behavior for large price movements. Therefore, it is an example of leptokurtic 
distribution, which means that extreme events occur more often than 
predicted by the Gaussian statistics. There is a large consensus that 
the empirical returns in the tails behave as a power law $p (x) \propto 1 / |x|^{1+\alpha}$, with a 
tail exponent $\alpha$ close to three in the high-frequency
 limit~\cite{Lux1996,Gopikrishnan1998,Gopikrishnan1999,Gopikrishnan1999b,Rak2007,Pan2008}.
 This implies that the variance of returns is finite. 
Power-law behavior has been recently 
detected in the scaling properties of extreme 
price movements in Bitcoin markets as well~\cite{Begusic2018}, albeit with a power-law exponent 
suggesting that cryptocurrency returns exhibit heavier tails than stocks.

To capture the above properties, 
different distributions borrowed from probability and statistics, as well as from statistical 
physics, have been proposed in the literature. The most important examples 
of non-Gaussian models that provide a good fit to the data are the truncated L\'evy 
distribution~\cite{BPBook2003,Gupta1999,Couto2001,Matsushita2003,Mariani2007}~\footnote{Empirical studies of 
daily and intraday returns using pure 
L\'evy distributions can be found in the classical econophysics papers~\cite{Mantegna1991,MSN1995} 
and, more recently, in~\cite{Scalas2007,Alfonso2012,Liu2023}. }, 
the Student's $t$-distribution~\cite{BPBook2003,Praetz1972,Blattberg1974,Peiro1994,Platen2008,Gu2008,Gerig2009,Koning2018}, 
the $q$-Gaussian (Tsallis) distribution~\cite{Tsallis2003,Rak2007,Katz2013,Alonso2019}, 
the hyperbolic distribution~\cite{Eberlein1995,Kuchler1999} 
and the modified
Weibull or stretched exponential distribution~\cite{Sornette1998,Sornette2003,Sornette2005,Kotz2006}. 
 A comparison of distributions in fitting the tails of daily index fluctuations of several stock
 markets can be found in~\cite{Eryigit2009}.

The results of the empirical analyses are also compatible with the scenario where price returns 
on liquid markets are,
beyond a correlation time of a few tens of minutes, 
uncorrelated random variables~\cite{MSBook2000,BPBook2003,Cont2001} 
and where the log-return distribution
converges very slowly to a Gaussian 
by aggregation~\cite{MSBook2000,BPBook2003,Gopikrishnan1999,Gopikrishnan1999b}. 
However, it is also known that returns exhibit non-linear correlations between their absolute value or 
square~\cite{MSBook2000,BPBook2003,Cont2001,Jondeau2006} and, therefore, they can not be treated 
as independent variables. The latter feature
has to be ascribed to the return volatility.

A direct consequence of the empirical behavior of stock prices is that all those quantities and strategies 
that are intimately related to the return dynamics are strongly affected by the anomalous properties of
their statistics. Typical examples are given by the fair price of stock options, portfolio
management and evaluation of market risk measures. In all these contexts, the price changes
play the role of fluctuations associated to underlying risky assets.

According to the standard and still widespread approach used in quantitative finance, the stochastic 
motion of stock prices is modeled using a
Geometric Brownian 
Motion (GBM)~\cite{Osborne1959,Samuelson1965,Gardiner2009,Shreve2013}, which implies that 
log-returns are normally distributed. The rationale behind this assumption is that the simplicity of
the Gaussian modeling and, more importantly, the whole machinery of It{\^ o}
stochastic calculus can be fully exploited to obtain analytically tractable results. The most 
prominent example is given by the celebrated Black and Scholes 
model of option pricing~\cite{BS1973} (see also \cite{Merton1973}).

In recent times, the overwhelming evidence about the leptokurtic nature of stock returns 
stimulated new directions in scholarly research and financial practice. For example, 
in the financial industry today, the risk measures, such as Value-at-Risk and Expected Shortfall, are estimated by using historical 
data and computing them from the 
percentiles of the distribution of real data, instead of using a 
parametric approach based on the normal approximation. In finance and econophysics, 
non-Gaussian closed-form expressions for the 
risk measures~\cite{Heikkinen2002,Kamdem2005,Bormetti2007} have been 
obtained by  analytically modeling the fat-tailed nature of price changes. 
Concerning option pricing, in the model proposed by Heston in finance~\cite{Heston1993},  
 the return volatility follows a random process and provides a more accurate option pricing than in the
Black and Scholes model. 
In finance and econophysics, 
many papers addressed the problem of non-Gaussian option 
pricing~\cite{Boyarchenko2000,Matacz2000,Borland2002a,Borland2002b,Kleneirt2002,Borland2004,Cassidy2010} 
using, in particular, L\'evy, Student's $t$ or Tsallis distributions to model the stock price dynamics.

Given the above motivations, the aim of this paper is to perform a systematic comparison of the main 
non-Gaussian models of the financial market dynamics. To characterize the probability 
of extreme price fluctuations, the deviations from the Gaussian behavior are parametrized 
in terms of tail exponents as obtained in the empirical studies of 
high-frequency returns. By using different Monte Carlo (MC) algorithms, we generate 
large samples of random deviates as synthetic data representations of financial
fluctuations, in order to compare their statistical properties and simulate their 
dynamical evolution. To the best of our knowledge, a comparative study of this kind 
represents an original addition to the literature and may
contribute to elucidate the similarities and universal properties of the existing models. 
It also opens the way to a number of possible applications.
As a first example, we apply our modeling to option pricing and study the differences 
with the results of the standard Gaussian approach {\it \`a la} Black and Scholes. 
 
 The paper is organized as follows. In Section~\ref{tf} we review the theoretical models 
 and introduce the probability distributions considered in our study. In Section~\ref{rd} 
 we present and discuss the results of our simulations, focusing on the behavior of the models
 in the tails, the dynamics of non-Gaussian asset prices and 
 the convergence rate to the asymptotic distributions. In Section~\ref{op}
 we investigate the implications of our modeling for option pricing. The main conclusions and 
 perspectives of the work are drawn in Section~\ref{cp}.

\section{Theoretical Modeling of Return Distributions}
\label{tf}

The asset price models used in our study are those that received the most 
attention in finance and econophysics in recent years. In agreement with the empirical findings, all of them 
are specified by probability density functions (PDFs) with zero mean (as we are interested 
to model price changes with subtracted average returns), finite variance and a positive excess kurtosis 
as a measure of outliers. For each PDF, the shape parameters are chosen to fit at best 
empirical data in the tails, as detailed below. 
Also note that we consider symmetric PDF as 
it is known that empirically skewness effects in the high-frequency return distributions are 
quite small, much smaller than the large kurtosis contribution~\cite{BPBook2003,Eryigit2009,Peiro1999,Tsay2010}.

We list in the following the main formulae defining the distributions that are considered in our study. 
In the equations below, the real-valued variable $x$ 
can be equally understood as return or log-return.
\vskip 8pt

$a)$ Student's $t$-distribution
\vskip 8pt
It has been widely used to successfully model returns both in finance and econophysics. The 
(generalized or scaled) Student's $t$, with zero location parameter, has PDF given by
\begin{equation}
p_S (x) \, = \, \frac{\Gamma(\frac{\nu+1}{2})}{\Gamma(\frac{\nu}{2}) \, \sqrt{\pi \nu} \hat{\sigma}} 
\left[  1 +  \frac{1}{\nu} \left( \frac{x}{\hat{\sigma}} \right)^{2} \right]^{-(\nu + 1)/2} \, ,
\label{eq:ps}
\end{equation}
where $\hat{\sigma}$ is a scale parameter and $\nu$ is the number of degrees of freedom that plays the role of 
shape parameter. In Eq.~(\ref{eq:ps}), the symbol $\Gamma$ denotes the Gamma function.
 For $\nu > 2$, the variance is $\hat{\sigma}^2 \nu / (\nu -2)$; the excess kurtosis is 
$6 / (\nu - 4)$ for $\nu > 4$ and infinite for $2 < \nu \leq 4$. In the asymptotic limit $|x| \to \infty$, the 
Student's $t$ behaves as a power law of the form 
\begin{equation}
p_S (x) \xrightarrow[ {|x| \to \infty} ] {} |x|^{-(\nu + 1)} \, .
\end{equation}
The typical values for $\nu$ found in the empirical studies are $\nu \simeq 3$ for 
intraday returns~\cite{BPBook2003,Gu2008,Gerig2009} 
and $\nu \simeq 4$ for daily returns~\cite{BPBook2003,Praetz1972,Platen2008,Koning2018}.
\vskip 8pt
$b)$ $q$-Gaussian distribution
\vskip 8pt
This distribution arises from Tsallis non-extensive statistical mechanics and has been applied in many
different disciplinary contexts, including modeling of financial data. Its standard PDF is given by
\begin{equation}
p_q (x) \, = \, \frac{\sqrt{\beta}}{C_q} \, e_q (- \beta x^2) \, ,
\label{eq:pq}
\end{equation}
where $\beta$ is a scale parameter, $C_q$ is a normalization factor and $e_q (\bullet)$ is the 
$q$-exponential function given by ($q \neq 1$)
\begin{equation}
e_q (x) \, = \, [1 + (1 - q) x]^{1 / (1-q)} \, .
\end{equation}
For $1 < q < 3$, which is the range of our concern, the normalization constant reads as follows
\begin{equation}
C_q = \sqrt{\frac{\pi}{q-1}} \, \frac{\Gamma \left( \frac{3-q}{2 (q-1)} \right)}{\Gamma \left( \frac{1}{q-1} \right)} \, .
\end{equation}
For $q < 5/3$ the variance is $\beta^{-1} / (5 - 3q)$; the excess kurtosis exists for $q < 7/5$ and is given by
$6 (q - 1) / (7 - 5q)$.
The shape of the PDF is determined by the $q$ parameter ($q \to 1$ yielding the normal 
distribution) and for $q > 1$ the $q$-Gaussian has asymptotic 
heavy tails given by the power law
\begin{equation}
p_q (x) \xrightarrow[ {|x| \to \infty} ] {} |x|^{- 2 / (q - 1)} \, .
\end{equation}

Note that there is a direct mapping between the Student's $t$ and $q$-Gaussian distribution. Actually, given a 
$q$-Gaussian with parameter $q$, the equivalent Student's $t$, as given by Eq.~(\ref{eq:ps}), is obtained by 
applying the following replacements in Eq.~(\ref{eq:pq})
\begin{equation}
q = \frac{\nu + 3}{\nu +1} \qquad {\rm with} \qquad  \beta = \frac{1}{(3-q) \hat{\sigma}^2} \, \, .
\end{equation}
According to the available econophysics studies for intraday and daily returns, the tail parameter $q$ 
preferably lies in the range $q \simeq 1.4 \div 1.5$~\cite{Tsallis2003,Rak2007,Katz2013}.
\vskip 8pt
$c)$ Truncated L\'evy distribution
\vskip 8pt
For this popular model of asset price fluctuations, there is no analytical expression for the PDF 
but its characteristic function is known in closed form. Following the literature~\cite{BPBook2003,Couto2001,Mariani2007}, 
we consider a symmetric 
Truncated L\'evy Distribution (TLD) for which a power law is only valid in an intermediate range 
and decays exponentially beyond it. To account for this smooth exponential cut-off for large arguments, 
we adopt the expression for the characteristic function first proposed in \cite{Koponen1995}. 
It explicitly reads as follows
\begin{equation}
\phi_{\rm TLD} (k) \, = \, \exp \left[ -\gamma \, \frac{(k^2 + \lambda^2)^{\alpha/2} \cos [\alpha \arctan (|k|/\lambda) ]
 - \lambda^\alpha}{\cos\left( \frac{\pi}{2} \alpha \right)}  \right] \quad \, \, 
\alpha \neq 1 \, ,
\label{eq:ftld}
\end{equation}
where $\gamma$ is a scale parameter, $\lambda$ is a truncation parameter and $\alpha$ is the L\'evy characteristic exponent, 
with $0 < \alpha \leq 2$, 
but $\alpha \neq 1$.
For $\lambda \to 0$, Eq.~(\ref{eq:ftld}) 
reduces to the well-known expression of the characteristic 
function of a (symmetric) $\alpha$-stable distribution~\cite{MSBook2000,BPBook2003}, 
where $\alpha = 1$ and $\alpha = 2$ correspond to the Cauchy and Gaussian distribution, respectively.
Note that the TLD, in contrast to the two-parameter 
distributions discussed above and in the following, is defined in terms of three free parameters 
and that its shape is determined
by the parameters $\alpha$ and $\lambda$.

The associated PDF is obtained from the characteristic function through an inverse Fourier transform
\begin{equation}
p_{\rm TLD} (x) \,  = \, \frac{1}{2\pi} \int dk \, e^{- i k x} \phi_{\rm TLD} (k) \, ,
\label{eq:tldf}
\end{equation}
which is automatically normalized since $\phi_{\rm TLD} (k = 0) = 1$.  The truncated L\'evy flight 
 that one obtains by inserting Eq.~(\ref{eq:ftld}) into Eq.~(\ref{eq:tldf}) behaves
 as a power law  smoothed 
 by an exponential of the form  
$|x|^{-(1+\alpha)} \, e^{-\lambda |x|}$~\cite{Koponen1995}.

The first cumulants of the TLD can be computed as successive derivatives of the logarithm of its characteristic 
function and their expression can be found in~\cite{BPBook2003} 
for any $\alpha$ in the range $1 < \alpha \leq 2$. The typical value present in fitting data 
with a TLD~\cite{BPBook2003,Couto2001} 
or close to the characteristic 
exponent found in fits of return time series using a pure L\'evy distribution~\cite{MSN1995,Scalas2007,Alfonso2012,Liu2023}
is $\alpha = 3/2$. In this case, the variance and excess kurtosis are given by
\begin{equation}
\sigma^2_{\rm TLD} \, = \, \frac{3}{2 \sqrt{2}} \, \frac{\gamma}{\sqrt{\lambda}} \qquad 
\qquad k_{\rm TLD} \, = \, 
\frac{\sqrt{2}}{2} \, \frac{1}{\gamma \, \lambda^{3/2}} \qquad {\rm for} \, \, \, \, \alpha = \frac{3}{2} \, \, .
\label{eq:tldm}
\end{equation}
\vskip 8pt
$d)$ Modified Weibull distribution
\vskip 8pt
This model has been proposed in econophysics 
in \cite{Sornette1998,Sornette2003,Sornette2005} (see references in~\cite{Kotz2006} for its use in finance), in
 particular to describe the 
behavior of empirical log-returns for large variations in the tails.

The PDF of the Modified Weibull Distribution (MWD) is given by 
\begin{equation}
p_{\rm MWD} (x) \, = \, \frac{1}{2 \sqrt{\pi}} \, \frac{c}{\chi} \left(  \frac{|x|}{\chi} \right)^{c/2 - 1} \, e^{- \left( |x| / \chi \right)^c} \, ,
\end{equation}
where $\chi$ and $c$ are a scale and shape parameter, respectively. It can be seen as a particular case of a 
generalized Gamma distribution but with domain $x \in \mathbb{R}$. It reduces to a standard
 Gaussian PDF 
for $c = 2$ and $\chi = \sqrt{2}$. When the exponent $c$ is smaller than one, the PDF is 
characterized by the presence of a stretched exponential, 
which decays more slowly than an exponential function and is found 
in various systems in nature and society~\cite{Sornette1998}.

The variance and excess kurtosis of the MWD are given by
\begin{equation}
\sigma^2_{\rm MWD} \, = \, \chi^2 \, \frac{\Gamma \left( \frac{1}{2} + \frac{2}{c} \right)}{\sqrt{\pi}} \qquad \qquad 
k_{\rm MWD} \, = \, \frac{\Gamma \left( \frac{1}{2} + \frac{4}{c} \right)}{\left[\Gamma \left( \frac{1}{2} + \frac{2}{c} \right) \right]^2} \, \sqrt{\pi} - 3 \, \, .
\label{eq:mwdm}
\end{equation}
The best fit values for the shape parameter $c$ mostly found in the empirical studies are 
in the range $c \simeq 0.6 \div 0.9$~\cite{Sornette1998,Sornette2005}.

\begin{figure}[h]
\centering
\includegraphics[width=12cm]{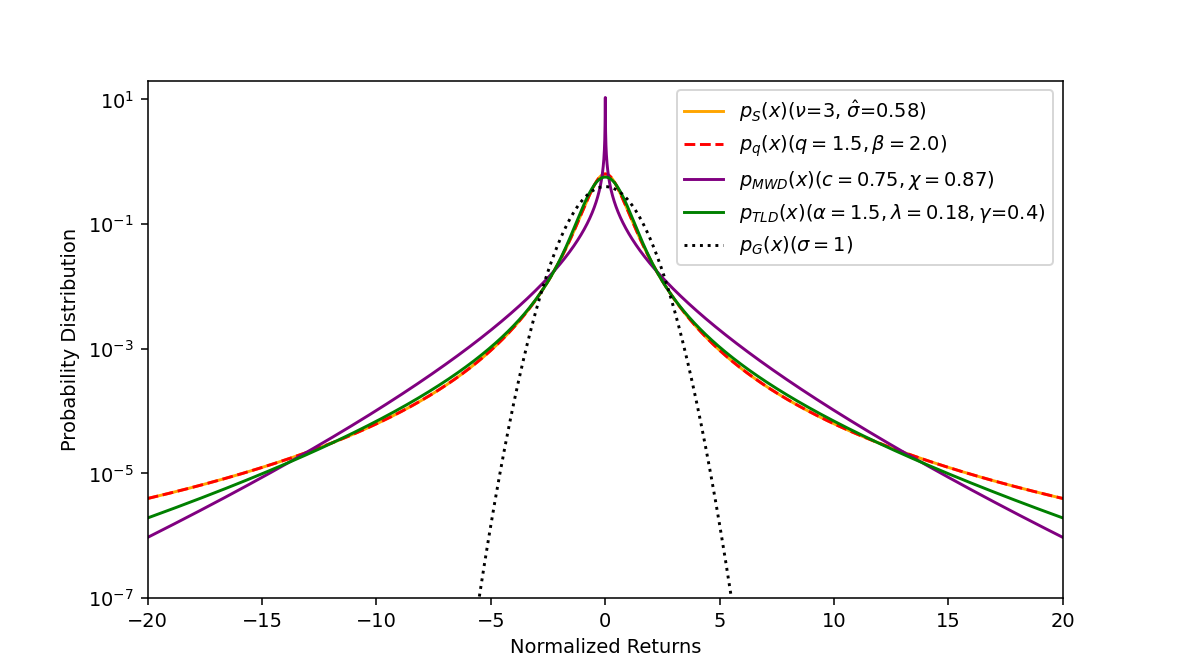}
\caption{Comparison between standardized (zero mean and unit variance) 
non-Gaussian return distributions and the Gaussian (dotted line) 
for normalized returns in the interval [-20,+20]. The choice 
of the distribution parameters used to model the behavior of high-frequency 
normalized returns is 
explained in the text. Note the log scale to emphasize the differences with the Gaussian shape and the similarities
of the non-Gaussian distributions along the tails when using realistic and consistent shape parameters.}
\label{fig:fig1}
\end{figure}

In Fig.~\ref{fig:fig1}, we show a first comparison between the aforementioned non-Gaussian 
distributions. For 
consistency, we normalize all the PDFs to 
have unit variance according to the following procedure.~\footnote{Note that 
the PDF standardization agrees with the strategy adopted in many of the fits to financial data, 
where returns are normalized by subtracting the empirical mean and dividing by the measured
standard deviation.} For the two-parameter distributions, 
we first fix the shape parameter 
by using one of the preferred values found in the empirical analyses of intraday returns: 
$\nu = 3$ for the Student's $t$, $q = 1.5$ for the $q$-Gaussian and $c = 0.75$ for the MWD. 
Then, the scale parameters, i.e. $\hat{\sigma}$ (Student's $t$), $\beta$ ($q$-Gaussian) and 
$\chi$ (MWD), are chosen such that the variance is equal to one for each distribution. For the 
three-parameter TLD, we first impose $\alpha = 3/2$ and then use Eq.~(\ref{eq:tldm}) 
to derive the scale parameter $\gamma$ as $\gamma = 2 \, \sqrt{2} \, \sqrt{\lambda} \, / \, 3$, that ensures 
unit variance. The truncation parameter $\lambda$ is finally chosen such that 
the two-parameter $\lambda$ and $\gamma$ 
theoretical combination used in fitting high-frequency data agrees with the measured value~\cite{BPBook2003}. 
In so doing, we obtain $\lambda \simeq 0.18$ (and thus $\gamma \simeq 0.4$ from the relation above), 
as better explained in the following. Our overall strategy closely 
follows the procedure used in~\cite{BPBook2003} to constrain the free distribution
parameters when fitting empirical data. In Fig.~\ref{fig:fig1}, the range for the normalized returns 
is chosen to be $[-20,+20]$ to mimic the outliers with largely positive kurtosis observed in 
the markets in the high-frequency limit~\cite{MSBook2000,BPBook2003,Cont2001,Jondeau2006,MSN1995,Gu2008,Tsallis2003}.

It is worth noting that, according to our empirically motivated choice of the 
free distribution parameters, the kurtosis of the Student's $t$ and $q$-Gaussian 
PDFs does not exist, whereas the kurtosis of the TLD and MWD distributions is 
large but finite. This is due to the fact that the 
non-Gaussian models under 
consideration belong to two different classes of distributions. Indeed, the
Student's $t$ and $q$-Gaussian PDFs follow asymptotically a pure power law
that implies an infinite kurtosis. On the contrary, for the TLD and MWD 
distributions, the power law behavior is smoothed by an exponential or 
stretched exponential decay that yields a finite kurtosis. This issue is 
further discussed in the following and in \ref{kur}.
   
As can be seen in Fig.~\ref{fig:fig1}, the similarity between the considered non-Gaussian models is
notable, with a comparable leptokurtic behavior and 
a quite good agreement along the tails, where extreme events model the large price movements 
occurring in financial markets. In particular, the common behavior of the 
 Student's $t$ and $q$-Gaussian modeling is due to the equivalence of the 
two parameterizations under our choice of the tail parameters, as discussed above.
More generally, the similar features of the non-Gaussian return 
distributions are further scrutinized and motivated in the 
next Section.

\section{Monte Carlo Simulations and Non-Gaussian Dynamics}
\label{rd}

To investigate the statistical properties of the non-Gaussian return models and 
simulate their dynamics, it is necessary having at hands MC samples 
of pseudorandom numbers (shortly denoted as random numbers or random deviates in 
the following) drawn from the distributions under study. This topic is 
addressed in the present Section, along with a study of the
convergence to the asymptotic distributions of the non-Gaussian stochastic processes. 

\subsection{Generation of Non-Gaussian Random Deviates and Tail Behavior}

A number of MC algorithms have been used and cross-checked to generate non-Gaussian 
random deviates as synthetic data representations of actual financial fluctuations. 

Generally speaking, for the Student's $t$, $q$-Gaussian and MWD models, whose 
PDFs are known analytically in direct $x$-space, a standard acceptance-rejection algorithm has been 
adopted to generate the associated random numbers. For the $q$-Gaussian, the results 
have been successfully cross-checked with those obtained by using the generalized 
Box-Muller method of~\cite{Thistleton2007,Nelson2021}. Note that a simple variation of the latter algorithm 
can provide random numbers for a Student's $t$ distribution as well, hence we used it 
also to verify the results of the acceptance-rejection method for this distribution, finding agreement. 

The random deviates associated to the TLD have been generated according to the
following step-by-step
procedure. We first computed the inverse Fourier transform as in Eq.~(\ref{eq:tldf}) 
by using the Fast Fourier Transform algorithm and then interpolated the resulting points 
with a cubic spline interpolation, in order to obtain a smooth $p_{\rm TLD} (x)$ function 
in the direct space and lastly apply an acceptance-rejection algorithm. 
To test the overall strategy, we checked that in the limit $\lambda \to 0$ 
(i.e. $\lambda$ vanishingly small in the simulations), where 
 Eq.~(\ref{eq:ftld}) reduces to the characteristic function of a pure $\alpha$-stable distribution, 
 our method provides MC samples in agreement with those obtained with the algorithm 
 of~\cite{Chambers1976,Weron1996} for different values of $\alpha$ 
 (see~\cite{Mantegna1994} for an independent algorithm to simulate 
 L\'evy stable processes). In particular, we checked that Cauchy 
 and Gaussian random deviates are correctly recovered for $\alpha = 1$ and $\alpha =2$, respectively.

\begin{figure}[h]
\centering
\includegraphics[width=7cm]{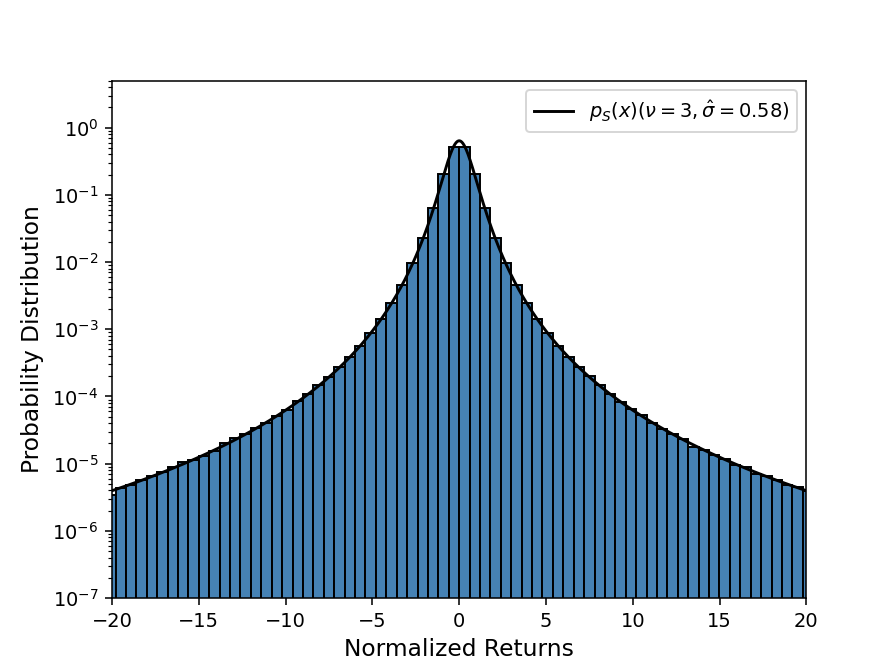}~~~~~\includegraphics[width=7cm]{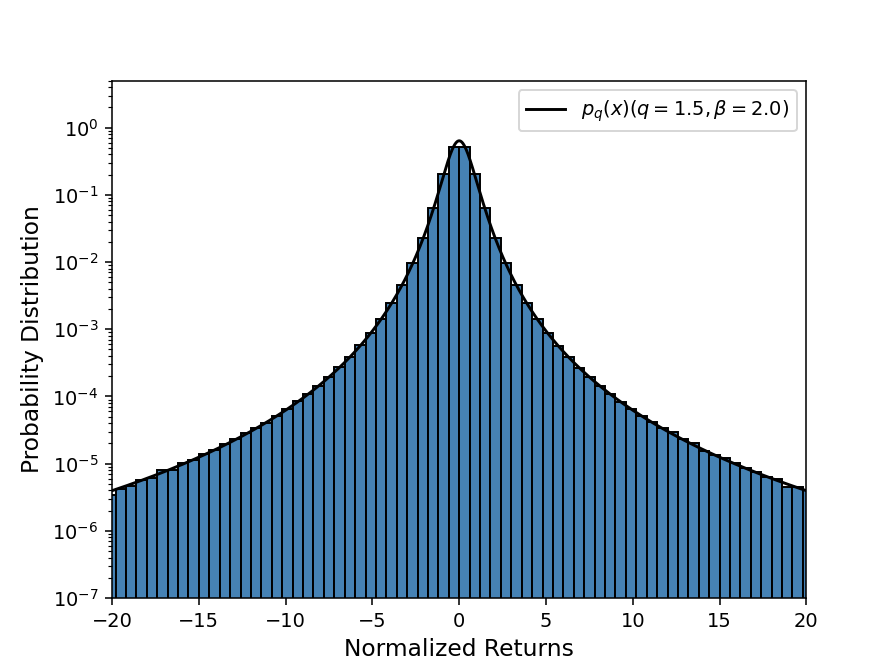}
\includegraphics[width=7cm]{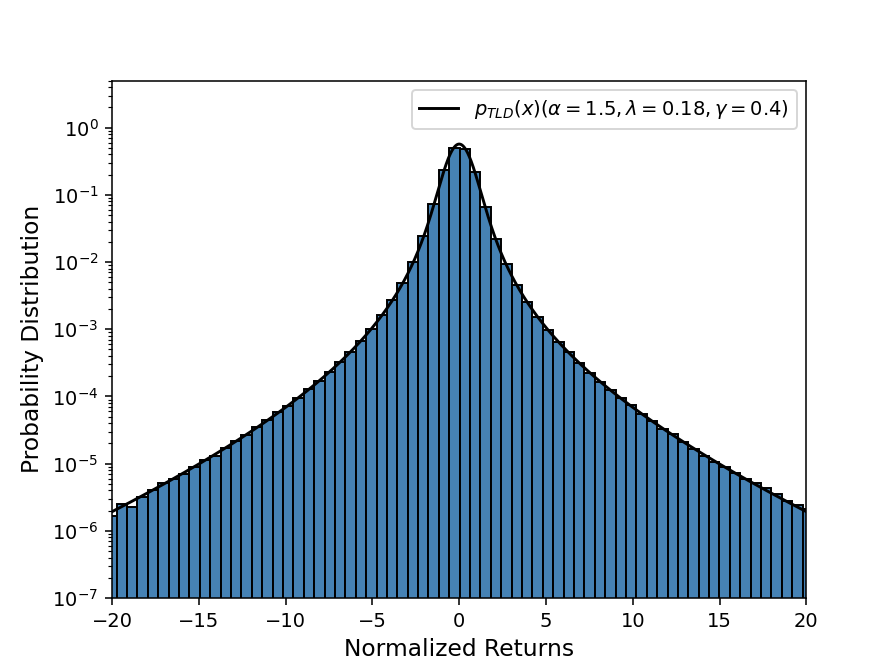}~~~~~\includegraphics[width=7cm]{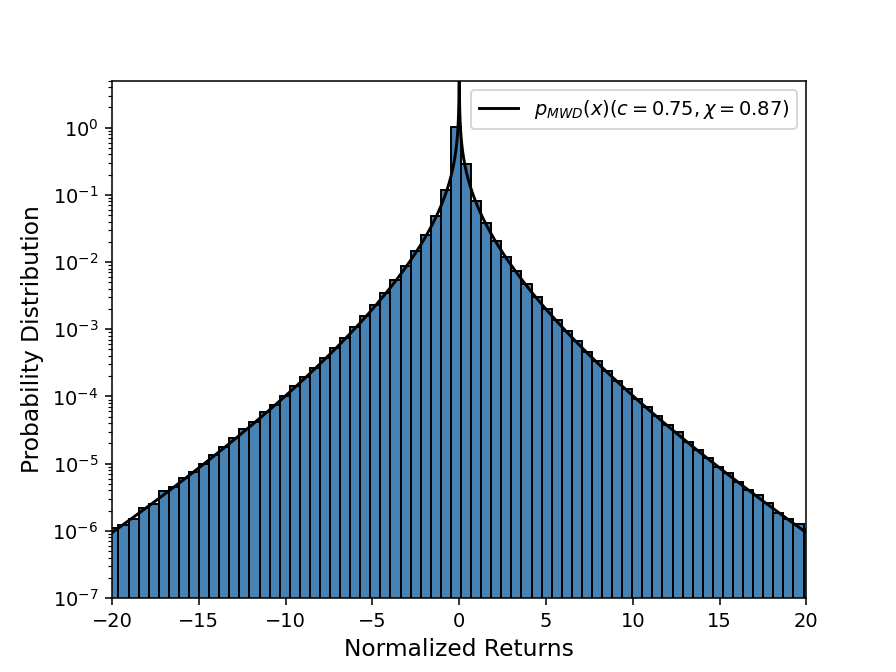}
\caption{Log scale comparison between the normalized non-Gaussian 
random number distributions (blue histograms) and the analytical expressions of the associated 
standardized PDF (solid line)
over the interval [-20,+20]. Upper panel: 
Student's $t$ ($\nu = 3$) and $q$-Gaussian ($q = 1.5$) distributions. Lower panel: TLD 
($\alpha = 3/2$, $\lambda = 0.18$) and MWD ($c = 0.75$) models.}
\label{fig:fig2}
\end{figure}

An example of the results of the above MC experiments is shown in Fig.~\ref{fig:fig2}. 
For each model, we generated $M = 10^8$ non-Gaussian 
random numbers, denoted by $\xi_{NG}$, by using the acceptance-rejection algorithm 
over the wide but bounded support [-30,+30], in order to sample the typical probability associated 
to the large jumps occurring in the markets that yield 
a substantial kurtosis~\cite{BPBook2003,Cont2001,Jondeau2006,Platen2008,Gu2008,Tsay2010,DelMar2012}. 
As can be 
seen from Fig.~\ref{fig:fig2}, the normalized histograms corresponding to the generated MC 
events nicely agree with the
analytical expression of the PDF for each non-Gaussian model. 

It is worth noting that, for the Student's $t$ and $q$-Gaussian distributions with 
ill-defined kurtosis, 
the generation of non-Gaussian random variables over a large but bounded domain
introduces an effective cutoff in their asymptotic power-law behavior. This makes 
the kurtosis of the associated random samples comparable to 
the empirical one and stable as a function of $M$, as shown in \ref{kur}. Actually,
 in the financial markets, the returns 
do not vary in their whole unlimited range 
and the measured kurtosis is typically large but obviously
limited, even in the presence of empirical power-law tails~\cite{DelMar2012}.  
Hence, our procedure resembles the sharp truncation of a pure $\alpha$-stable distribution 
to get a truncated L\'evy flight
 with finite variance and higher moments~\cite{Mantegna1994b}, that can be used as a 
 model of the price change statistics~\cite{MSBook2000,Gleria2002,Figueiredo2003,Grabchak2010,Schinckus2013}. 
 This approach is further supported by the good fit to extreme 
 price movements
 provided by a truncated Pareto model~\cite{Aban2006} and by the 
 decline of the kurtosis towards normality with time aggregation~\cite{Wu2006}.
  
  Note also
 that the generation of $\xi_{NG}$ over a finite range introduces a (small) bias in the characterization
  of their unit variance. For this reason, 
we normalize all the generated random deviates through the replacement $\xi_{NG} \to \xi_{NG} / s$, 
where $s$ is the sample standard deviation of each non-Gaussian MC sample~\footnote{We do not subtract the sample 
mean in the normalization 
procedure as its value is negligible, at the level of $10^{-4} \div 10^{-5}$, for all the models.}. 
According to this overall strategy, all our random numbers 
can be seen as synthetic representations of normalized financial returns, as further discussed 
in the following.

We used the above $M = 10^8$ random deviates to investigate the behavior of the non-Gaussian 
return models in the tails. To this end, we studied the complementary cumulative distribution 
function (CCDF) which, for a given PDF $p(x)$, is defined as 
\begin{equation}
P (x) \, = \, \int_{x}^{+\infty} p(x^{\prime}) \, d x^{\prime} \, .
\label{eq:ccdf}
\end{equation}
As it is well known, it gives the probability that the real-valued random variable under 
consideration will take a value greater than or equal to the threshold $x$. 

\begin{figure}[h]
\centering
\includegraphics[width=9.5cm]{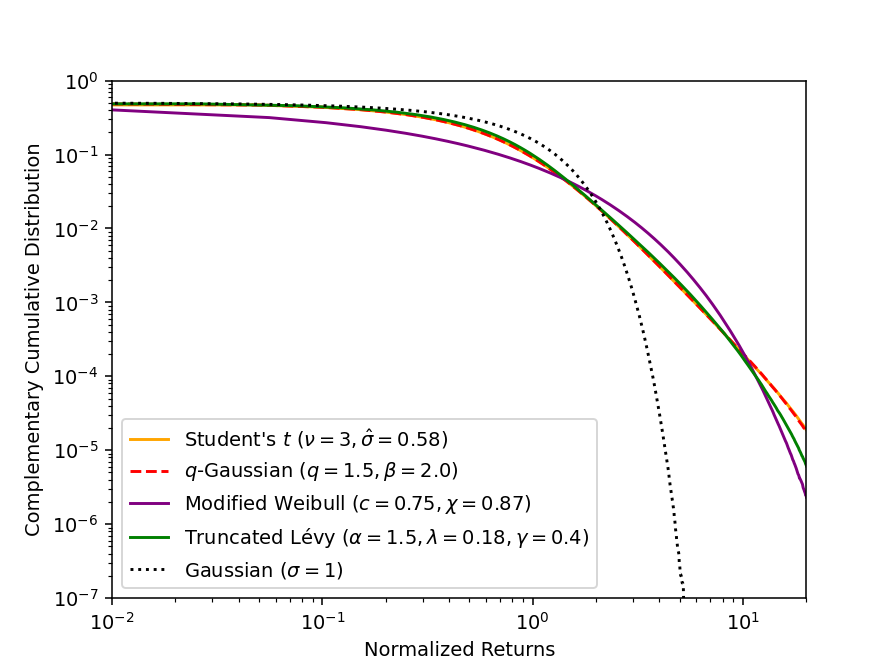}
\caption{Log-log scale comparison between the CCDFs of standardized non-Gaussian return distributions 
and the Gaussian CCDF (dotted line) for normalized returns 
above threshold in the range [0.01,20]. The distribution parameters 
modeling the high-frequency dynamics 
are the same as those
in Fig.~\ref{fig:fig1}. $M = 10^8$ non-Gaussian random deviates are generated over the 
interval [-30,+30] and used to compute 
each CCDF. 
}
\label{fig:fig3}
\end{figure}

For each non-Gaussian model, we computed the probability of Eq.~(\ref{eq:ccdf}) 
from the generated random number samples. The results of this comparative
analysis are shown in Fig.~\ref{fig:fig3} for normalized 
returns above threshold in the range [0.01,20]. As can be noted, 
the CCDFs associated to the different 
distributions not surprisingly display some difference but there is also
a noticeable overall consistency for what concerns the 
dynamics of extreme price changes along the tails. 

This substantial agreement
is not accidental and can be explained as follows. For the Student's $t$ and 
$q$-Gaussian distributions, the choice of the shape parameters $\nu = 3$ and 
$q = 1.5$ implies that both PDFs follow asymptotically the power law 
$p (x) \propto x^{-4}$ (for positive $x$) and therefore the asymptotic behavior 
of both CCDFs is given by $P(x) \propto x^{-3}$. This trend is known as
 {\it inverse cubic law}~\cite{Lux1996,Gopikrishnan1998,Gopikrishnan1999,Gopikrishnan1999b,Rak2007,Pan2008}, 
 which is considered to be a robust stylized fact as it is
  generally observed in the high-frequency regime for different 
stock markets and asset prices. On the other hand, 
 the Student's $t$  and TLD provide similarly good fits to financial
fluctuations in the tails, as shown in~\cite{BPBook2003}. For example, in a fit to the 
30 minute data of the S\&P 500 index, $\nu \simeq 3$ for the Student's $t$ distribution
is found to correspond for the $\alpha = 3/2$ TLD to the fitted quantity 
$\gamma^{2/3} \, \lambda = 0.096$~\cite{BPBook2003}. 
For our TLD parameter choice, i.e. $\lambda = 0.18$ and thus $\gamma = 0.4$ 
(from the unit variance constraint), we get 
the theoretical value $\gamma^{2/3} \, \lambda \simeq 0.098$, in good agreement with 
the measured parameter combination. 
As remarked in~\cite{BPBook2003}, the above quantity represents 
the ratio of the typical scale of the core 
of the TLD given by 
$\gamma^{2/3}$ to the scale $\lambda^{-1}$ 
at which the exponential cut-off takes place. 
Concerning the MWD model, it is known that the associated CCDF for the 
shape parameter $c$ in the range $c \simeq 0.6 \div 0.9$ ($c = 0.75$ in our case)
accurately describes extreme price movements of traded 
currency exchange rates and that, in general, one cannot clearly distinguish between a 
power law tail and a stretched exponential decay~\cite{Sornette2003}. As shown 
in~\cite{Sornette2003}, this is due to the fact that
the stretched exponential tends 
to the Pareto distribution in a certain limit where the shape parameter $c$ 
goes to zero.

To assess the reliability of our non-Gaussian modeling along the tails, we compared 
the kurtosis as computed from the generated MC samples with the theoretical results, whenever 
possible, as well as with the typically measured values. This cross-check is described in~\ref{kur}.

It is worth noting that a similar picture holds for a modeling of the thinner
 tails of daily returns with less pronounced kurtosis, when making an appropriate choice of the free distribution parameters.
Actually, we observed that the considered non-Gaussian models provide a similar 
description of 
price variations when using the following set of parameters in the standardized distributions: 
$\nu = 4$ for the Student's $t$,
$q = 1.4$ for the $q$-Gaussian, $\lambda = 0.26$ for the $\alpha = 3/2$ TLD and 
$c = 0.85$ for the MWD. 

To summarize, the similarities between the non-Gaussian return models as observed in 
Fig.~\ref{fig:fig1} and Fig.~\ref{fig:fig3} are a direct and natural consequence of a 
coherent and realistic choice of the distribution parameters, as well as of a large body 
of empirical evidences about the universal scaling properties of large price movements.
 
\subsection{Simulations of Non-Gaussian Return Dynamics}

The random numbers drawn from the distributions under consideration can be also 
used to simulate the dynamics of returns and log-returns under non-Gaussian models,
as shown in the following.

According to the standard model of finance, the time evolution of asset prices 
is described by a GBM given by the following stochastic differential equation (SDE)
\begin{equation}
\D S (t) \, = \, \mu \, S(t) \, \D t + \sigma \, S(t) \, \D W (t) \, ,
\label{eq:gbm}
\end{equation}
where $\mu$ is the percentage drift, $\sigma$ the volatility and $d W(t)$ 
the increment of a Wiener process (It{\^ o} prescription is assumed). 
Equation~\ref{eq:gbm} models the stochastic motion of 
returns, i.e. percentage price variations. By It{\^ o} lemma, it follows that the 
dynamics of log-returns is given by the SDE of a Brownian motion with drift, i.e.
\begin{equation}
\D \ln S (t) \, = \, \left( \mu \, - \frac{\sigma^2}{2} \right) \, \D t + \sigma \, \D W (t) \, .
\label{eq:abm}
\end{equation}
By integrating the above equations with the use of It\^o calculus~\cite{Gardiner2009,Shreve2013}, it turns out 
that log-returns follow a normal
distribution (with a variance growing linearly as a 
function of time, because of the properties of the Wiener process) and asset prices 
are log-normally distributed random variables.

Both Eq.~(\ref{eq:gbm}) and Eq.~(\ref{eq:abm}) can be simulated through a 
random walk MC algorithm in discrete time by using the Euler-Maruyama method~\cite{Gardiner2009}
given the following recursive equations
\begin{eqnarray}
&& S (t + \Delta t) \, = \, S(t) + \, \mu \, S(t) \, \Delta t + \sigma \, S(t) \, \xi_{G} \, \sqrt{\Delta t}  \label{eq:sg} \, , \\
&& \ln S (t + \Delta t) \, = \, \ln S(t) \, + \, \left( \mu \, - \frac{\sigma^2}{2} \right) \, \Delta t + \sigma \, \xi_{G} \, \sqrt{\Delta t} \, .
\label{eq:sgg}
\end{eqnarray}
In Eq.~(\ref{eq:sg}) and Eq.~(\ref{eq:sgg}), $\xi_{G}$ is a random number drawn from a standard normal 
distribution, i.e. $\xi_{G} \sim {\cal N} (0,1)$, and the discrete time increment $\Delta t$
has to be chosen sufficiently small, so that the numerical simulation provides a description as accurate as 
possible of the continuous-time processes given by Eq.~(\ref{eq:gbm}) and Eq.~(\ref{eq:abm}).

\begin{figure}[h]
\centering
\includegraphics[width=7.5cm]{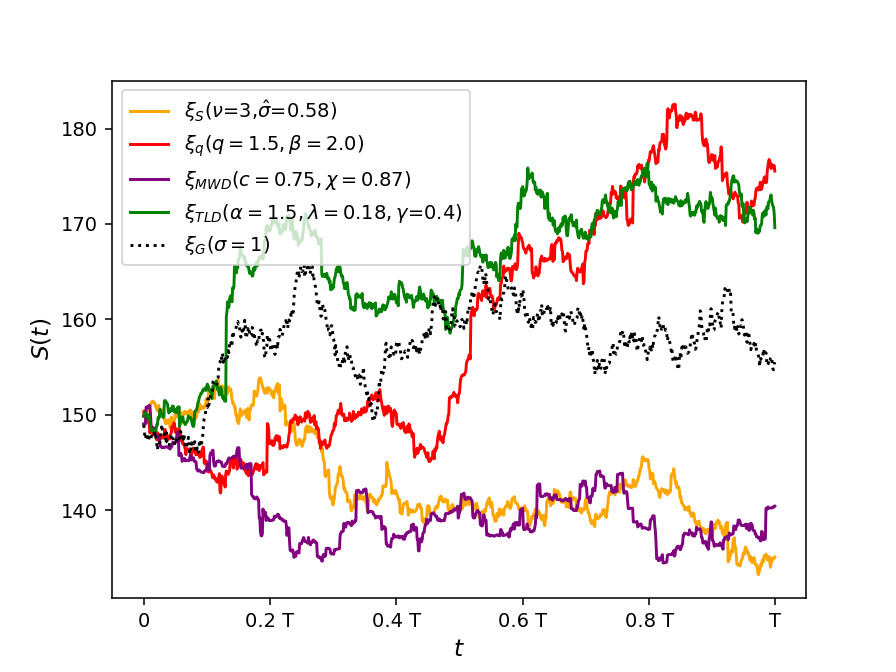}~~~~~\includegraphics[width=7.5cm]{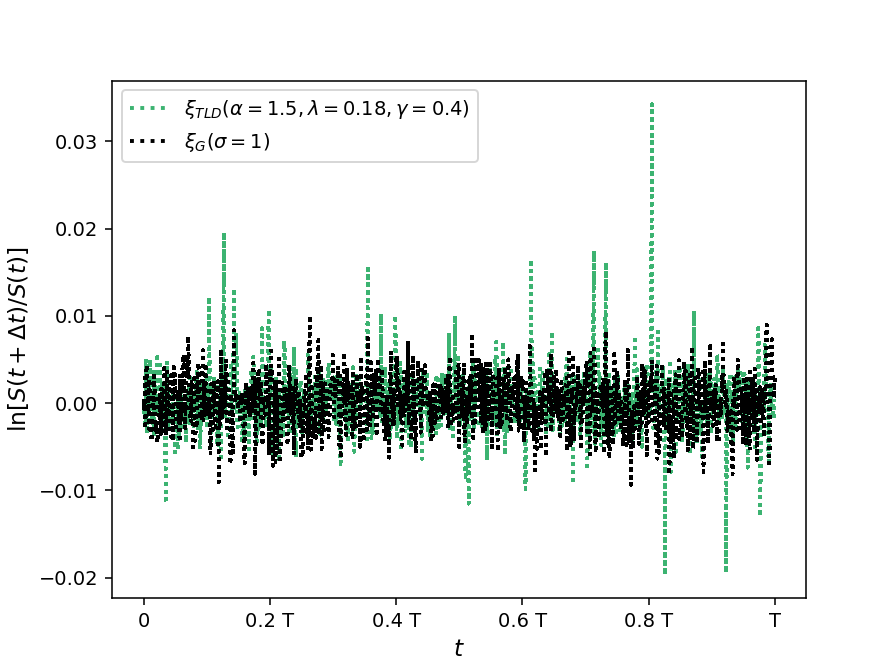}
\caption{Left panel: Non-Gaussian random walk simulations of asset price dynamics in comparison 
with the Gaussian modeling (black dotted line). The parameters of the standardized non-Gaussian 
distributions are the same as those of the previous Sections. Other quantities are: $\Delta t = 10^{-3}$,
$N = 10^3$ (number of iterations), so that $t = N \, \Delta t$ up to $T = 1$, $S(0) = 150$, 
$\mu = 0.01$ and $\sigma = 0.1$.
Right panel: log-returns 
for the TLD (green fluctuations) and Gaussian (black fluctuations) dynamics. 
Similar results 
hold for the other non-Gaussian models.}
\label{fig:fig4}
\end{figure}

Eq.~(\ref{eq:sg}) and Eq.~(\ref{eq:sgg}) can be generalized to non-Gaussian scenarios, 
through the random number replacement $\xi_G \to \xi_{NG}$, regardless of the considered 
non-Gaussian model. Note that, as previously remarked, the variables $\xi_{NG}$ are zero mean and unit variance
random numbers, thus ensuring consistency with the Gaussian 
formulation. It is also worth noting that in this way we are assuming that returns and 
log-returns in the presence of fat tails follow a standard diffusion process, namely with 
the same dependence on time of the variance of a Wiener process, and are treated as
statistically independent and identically distributed (i.i.d.) random variables. This strategy is followed
to compare the results of non-Gaussian random walks with those of Gaussian simulations.~\footnote{Actually, it is known that financial returns undergo a process 
of anomalous diffusion~\cite{Mantegna1991,MSN1995,Plerou2000,Michael2003,Alonso2019}, from the high-frequency regime 
to relatively long times. This feature 
can be implemented in our modeling as well and will be addressed in a future study. 
Also the assumption of independence could be relaxed, by treating the volatility as 
a stochastic process.}

In Fig.~\ref{fig:fig4} (left panel) we show the non-Gaussian random walk simulations of 
asset price dynamics 
for all the models under consideration, along with a comparison with the standard Gaussian
results. As expected, the non-Gaussian simulations display large jumps from time to time 
and therefore better resemble the observed dynamics of financial markets.
This feature is highlighted in the right panel of Fig.~\ref{fig:fig4}, where the log-return 
pattern of the TLD modeling is compared with the Gaussian scenario. 
The TLD dynamics is chosen for definitiveness only, as almost identical results 
hold for the other distributions.

\subsection{Convergence to the asymptotic distributions}

Before moving to the application of our modeling to
option pricing, we study the convergence to the
asymptotic distributions of the non-Gaussian stochastic processes.

 In the standard model of finance, log-returns 
are normally distributed at any time and asset prices always follow a log-normal 
distribution. These results can be derived by integrating the SDEs of Eq.~(\ref{eq:abm}) and 
Eq.~(\ref{eq:gbm}) under stochastic calculus~\cite{Gardiner2009,Shreve2013}. However, 
the emergence 
of the normal distribution from the additive stochastic process 
of Eq.~(\ref{eq:abm}) and of the log-normal 
distribution from the multiplicative random process of Eq.~(\ref{eq:gbm}) 
can be also explained as a consequence of the limit theorems of probability.
Since in our work the fat-tailed log-returns are modeled by 
i.i.d.~random variables with finite variance (and finite higher moments as well), one can expect that the MC simulations 
of non-Gaussian log-return dynamics 
converge to a normal distribution after a sufficiently large number of iterations, because
of the Central Limit Theorem (CLT)~\cite{MSBook2000,BPBook2003,Mantegna1994b,Taleb2020}. Similarly, the non-Gaussian 
asset price 
dynamics should converge asymptotically to a log-normal PDF, 
as a consequence of the Multiplicative Central Limit Theorem (MCLT)~\cite{Newman2005,Redner1990,Rempaia2002}. 
However, as largely discussed 
in the literature~\cite{MSBook2000,BPBook2003,Mantegna1994b,Romeo2003,Taleb2020,Draper2021},
the rate of convergence critically depends on the properties (i.e. the moments of 
order larger than two) of the elementary distribution 
describing the variables to be summed or multiplied. 

For definitiveness, we choose the TLD as model 
of non-Gaussian dynamics but the same general conclusions hold for the other distributions. 
We also examine the case of standard 
Gaussian dynamics as a cross-check of the results of It\^o calculus, as well as to
compare with the non-Gaussian simulations. Moreover, for the sake
of simplicity, but without loss of generality,  we 
 we use the relation $\mu = \sigma^2 / 2$ in Eq.~(\ref{eq:sg}) and Eq.~(\ref{eq:sgg}) 
 and we set the volatility $\sigma$ equal to one. 
We perform this convenient parameter choice as a standardization 
procedure in order to simplify the behavior of the 
limit distributions and their statistical moments, as the drift term 
(i.e. the average) in the log-return 
process vanishes and the instantaneous standard deviation $\sigma$ 
does not contribute.

Therefore, for the sum of i.i.d. random variables and the numerical test of the CLT, we 
consider the following additive stochastic processes in discrete time
\begin{eqnarray}
&&  \ln S (t + \Delta t) \, = \, \ln S(t) \, + \, \xi_{G} \sqrt{\Delta t}  \label{eq:cltg} \, , \\
&&  \ln S (t + \Delta t) \, = \,  \ln S(t) \, + \, \xi_{NG} \sqrt{\Delta t} \, ,
\label{eq:cltng}
\end{eqnarray}
where $\xi_G \sim {\cal N} (0,1)$ and $\xi_{NG} \sim {\rm TLD} (0, 1)$, i.e. $\xi_{NG}$
is a random number distributed according to a TLD with zero mean and unit 
variance, as by construction of our MC samples.

Correspondingly, by It\^o formula, the associated multiplicative stochastic processes in discrete time
are given by
\begin{eqnarray}
&& S (t + \Delta t) \, = \, S(t) +  \, \frac{S(t)}{2} \, \Delta t \, + \, S(t) \, \xi_{G} \sqrt{\Delta t}  \label{eq:mcltg} \, , \\
&& S (t + \Delta t) \, = \,   S(t) + \frac{S(t)}{2} \, \Delta t \, + \, S(t) \, \xi_{NG} \sqrt{\Delta t} \, ,
\label{eq:mcltng}
\end{eqnarray}
which are the equations used for the study of the MCLT.

\begin{figure}[h]
\centering
\includegraphics[width=7.cm]{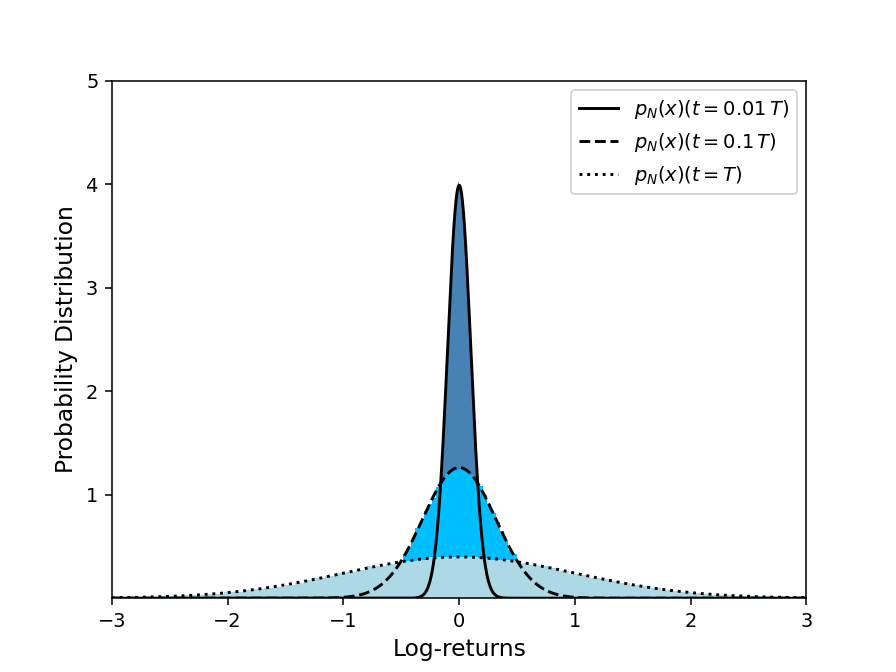}~~~~~\includegraphics[width=7.cm]{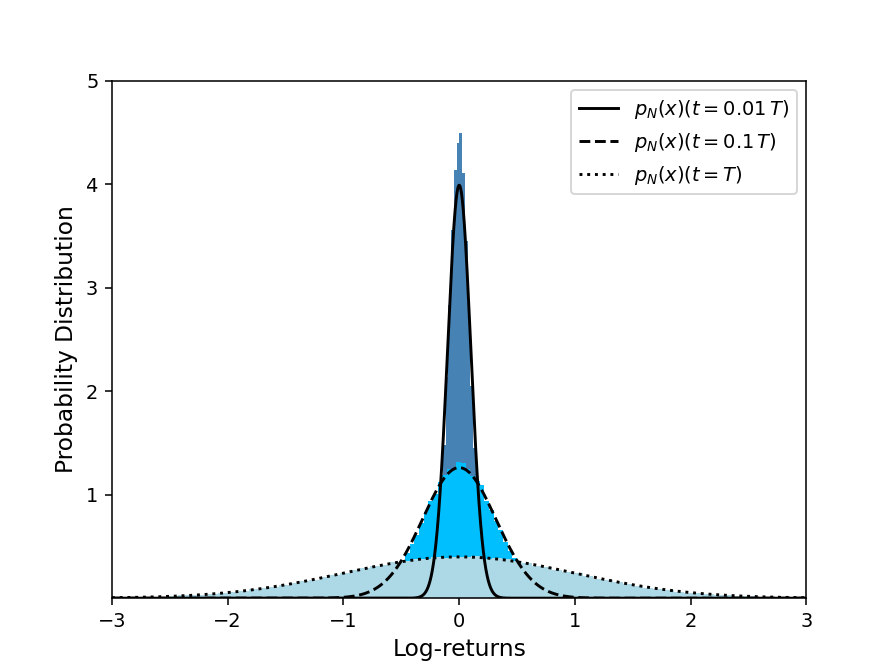}
\includegraphics[width=7.cm]{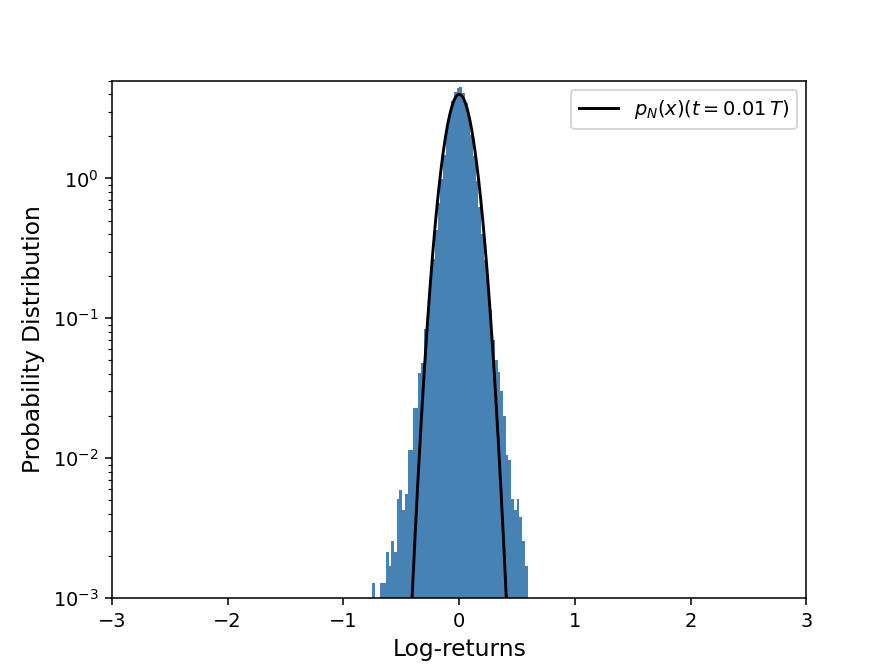}~~~~~\includegraphics[width=7.cm]{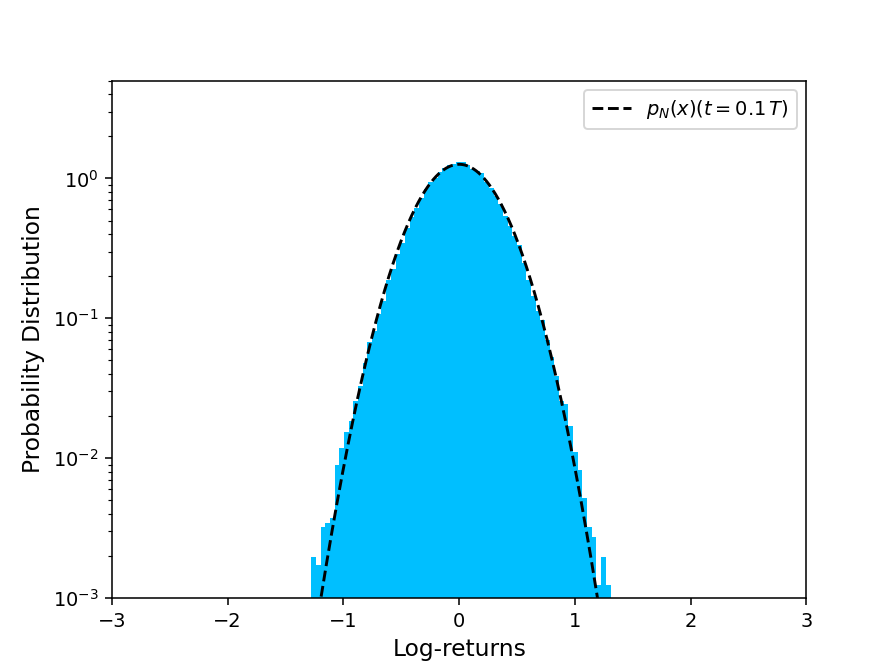}
\caption{Test of the CLT. Upper panel, linear scale: data distributions as a function of $N$ (colored histograms) 
for the sum of i.i.d. Gaussian (left plot) and TLD (right plot) random variables, in comparison with the
asymptotic normal distributions (lines). Lower panel, log scale: data distributions for the 
sum of $N = 10$ (left plot) and $N = 10^2$ (right plot) i.i.d. TLD variables in comparison with the asymptotic Gaussians (lines).}
\label{fig:fig5}
\end{figure}

In the above equations, we set $\Delta t  = 10^{-3}$ to approach the 
continuum limit of the actual dynamics and we investigate the convergence to the 
asymptotic distributions by summing or multiplying $N = 10, 10^2, 10^3$ i.i.d. variables. 
Hence, as noted in \cite{Mantegna1994b}, in our random walk simulations the 
variables $\xi_i$ ($i = G, NG$) are the jump sizes performed after a time interval 
$\Delta t$ and $N$ is the number of time intervals, so that the time elapsed 
after $N$ iterations is given by $t = N \Delta t$, i.e. $t = 0.01, 0.1, 1$, where 
$T = 1$ is the maximum time of the dynamical evolution. 
Therefore, in the 
figures of the present Section, the number of variables $N$ and the time $t = N \Delta t$ 
can be interchanged everywhere.

In our simulations, we use $S(0) = 1$ as initial condition of the dynamics.
 Following the literature, we compare our numerical results for both the moments and 
the shape of the limiting distributions, as a function of $N$, with the predictions
of the central limit theorems of probability. For the additive dynamics without drift 
of Eq.~(\ref{eq:cltg}) and Eq.~(\ref{eq:cltng}), the asymptotic distribution is a symmetric normal 
about the origin, whose first two non-zero raw moments are
$\mathbb{E} [\ln S(t)^2] = t$ and $\mathbb{E} [\ln S(t)^4] = 3 \, t^2$. For the 
multiplicative processes of Eq.~(\ref{eq:mcltg}) and Eq.~(\ref{eq:mcltng}),
the asymptotic distribution is a log-normal whose $n$-th raw moment grows exponentially as 
$\mathbb{E} [S(t)^n] = e^{n^2 t /2}$.

At each time $t = N \Delta t$, we compute the ensemble averages yielding the moments 
as sample averages over $M = 10^5$ realizations or trajectories. The data distributions 
are obtained as normalized histograms over the same samples. To account for
the finite size of the MC statistics, we associate confidence intervals 
to the calculation of the sample averages 
 by using the bootstrap resampling method~\cite{EfronBook1993}. 
In particular, we compute the two-sided bootstrap intervals at 99.7\% Confidence 
Level (CL) with the bias-corrected and accelerated (BCa) method~\cite{EfronBook1993},
as it provides more accurate intervals in the presence of high skewness or 
long tails in data distributions.

\begin{figure}[h]
\centering
\includegraphics[width=7.cm]{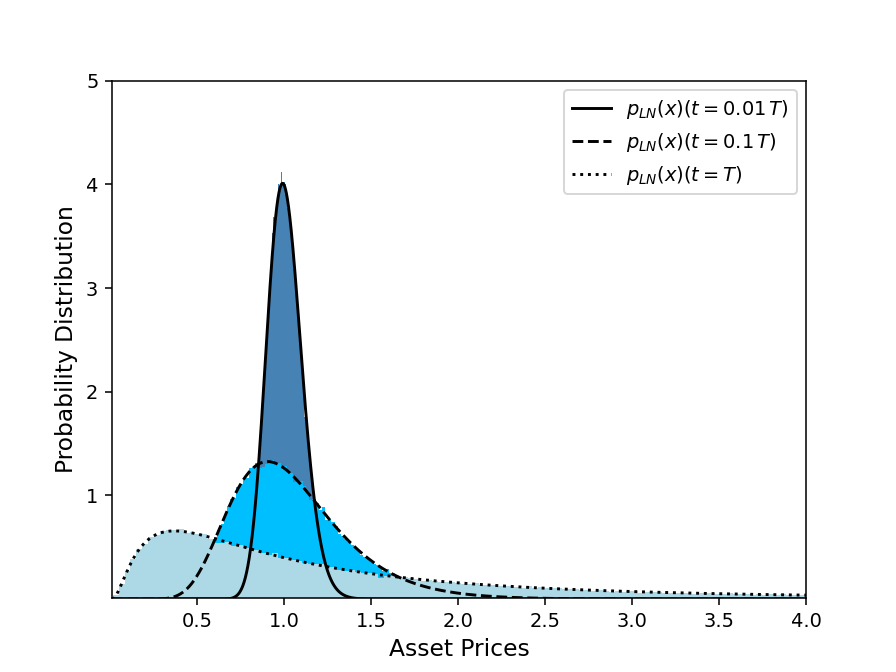}~~~~~\includegraphics[width=7.cm]{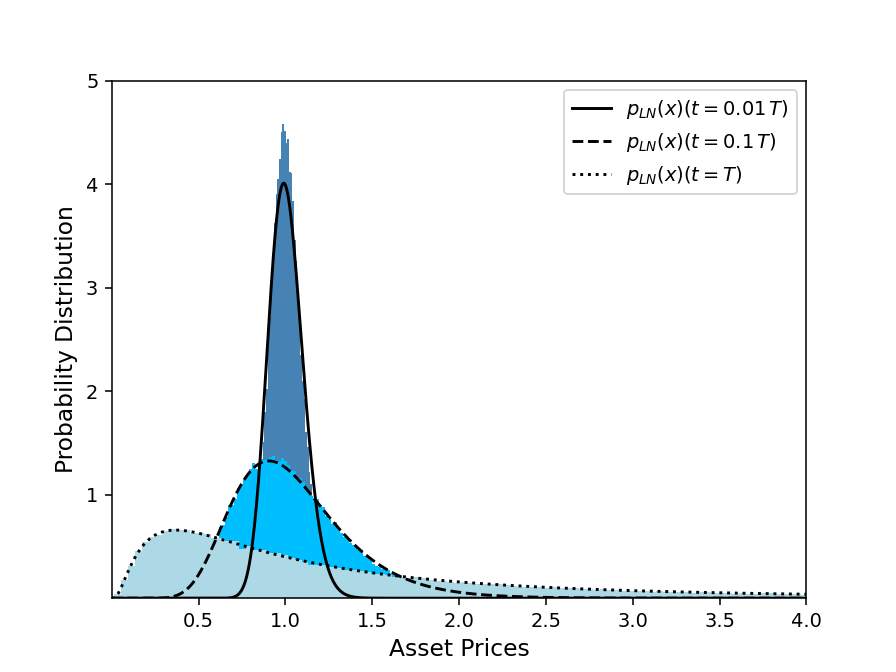}
\includegraphics[width=7.cm]{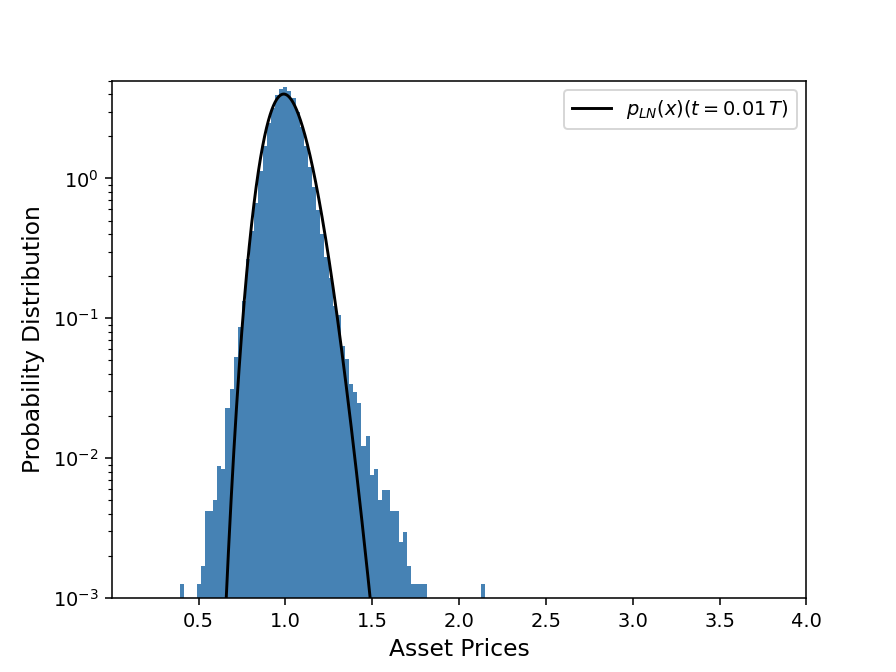}~~~~~\includegraphics[width=7.cm]{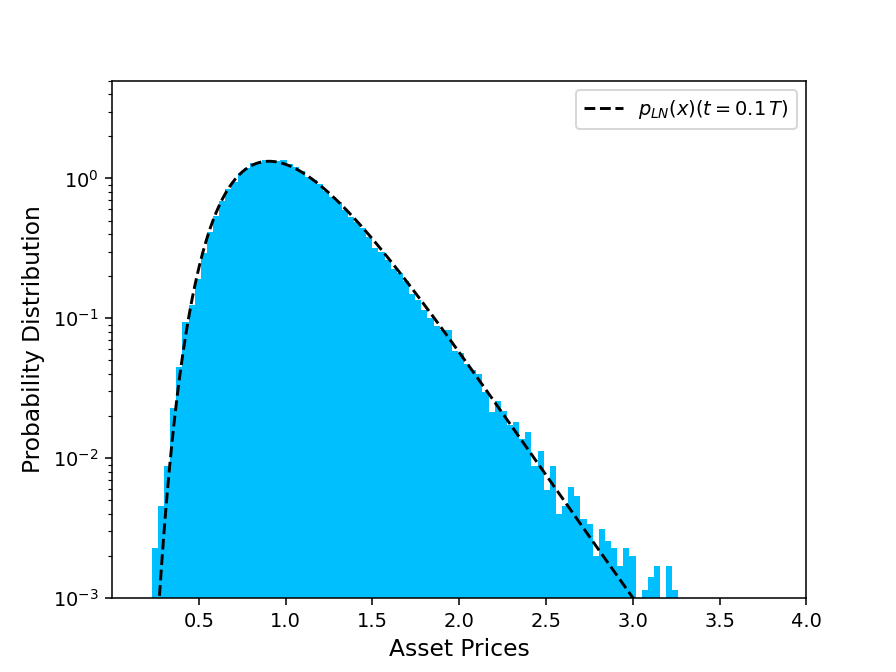}
\caption{Test of the MCLT. Upper panel, linear scale: data distributions as a function of $N$ (colored histograms) 
for the multiplication of i.i.d. Gaussian (left plot) and TLD (right plot) random variables, in comparison with the
asymptotic log-normal distributions (lines). Lower panel, log scale: data distributions for the 
multiplication of $N = 10$ (left plot) and $N = 10^2$ (right plot) i.i.d. TLD variables in comparison with the asymptotic 
log-normals (lines).}
\label{fig:fig6}
\end{figure}

The main results of this study are shown in Fig.~\ref{fig:fig5} for the CLT and in
 Fig.~\ref{fig:fig6} for the MCLT. 

As expected, for the sum of i.i.d. Gaussian variables, the convergence to the normal is immediately achieved, 
as we checked for the sample moments and it is evident from the data distributions for different 
$N$ in Fig.~\ref{fig:fig5} (upper panel, left plot). Of course, this is due to the stability property
of the normal distribution and is in agreement with It\^o calculus, that holds under the assumption of 
continuous-time Gaussian dynamics. For the sum of i.i.d. TLD variables, we observed in 
our simulations that the fourth raw sample moment, that measures the kurtosis of the 
numerically summed distributions, disagrees with the CLT expectation for $N = 10, 10^2$ 
(even at 99.7\% CL) 
and gradually converges to the theoretical prediction for some $N$ in the range $10^2 \div 10^3$. This feature can be 
appreciated from the shape of the data distributions in comparison with the 
asymptotic normals in Fig.~\ref{fig:fig5} (upper panel, right plot and lower panel). Following the remarks 
of~\cite{BPBook2003}, we checked that not only the central part but also the tails of the 
data distribution follow the normal shape when $N = 10^3$. Therefore, for the
TLD variables, the crossover to the Gaussian regime is quite slow as a function of 
the number of summands, since the speed of convergence to the asymptotic 
normal is ultimately controlled by the conditions on higher-order moments of the
 Berry-Esseen theorem and Chebyshev-Edgeworth expansion~\cite{MSBook2000,Feller1991,Draper2021}. Our findings 
 agree with the analytical considerations of~\cite{BPBook2003}, where, for the sum
 of $N$ i.i.d. TLD variables, the CLT is found to be valid under the inequality 
 $N \gg N^* = k_{\rm TLD}$, $k_{\rm TLD}$ being the kurtosis of the TLD. In our 
 modeling, $k_{\rm TLD} = 23.2$ (see \ref{kur}), that explains the observed low 
 convergence rate.
  
For the multiplication of i.i.d. Gaussian variables, we noticed 
that all the first four sample moments 
are compatible with the log-normal theoretical expectations of the MCLT for any $N$. 
We only observed some underestimate for the central value 
of the fourth sample moment at $N = 10^3$, with however a large 
BCa bootstrap CL, that signals a critical dependence of this ensemble 
estimate (the sample kurtosis of a broad distribution~\cite{Romeo2003}) from the number of 
realizations. Anyway, 
the data distributions nicely agree everywhere with the limiting log-normal PDFs for any $N$,
as can be seen from Fig.~\ref{fig:fig6} (upper panel, left plot).
Again, these results are consistent with the 
predictions of It\^o calculus about the asset price dynamics. It is worth noting that
our results do not display any deviation between the sample 
averages and the theoretical ensemble averages, that exhibit exponential 
growth as a consequence of the multiplicative nature of the process~\cite{Redner1990,Peters2013}. 
Actually, as remarked in \cite{Redner1990,Peters2013}, large outliers and atypical events 
for the sample size can lead
to a deviation from the correct ensemble-average behavior but only after 
a time $\tau \sim \ln (M)$, $M$ being the number of realizations. In our simulations, 
where $M = 10^5$, this critical time lies far beyond the maximum time of the 
dynamical evolution given by $T = 1$, corresponding to $N = 10^3$ multiplications 
with $\Delta t = 10^{-3}$. 

For the multiplication of i.i.d. TLD random numbers, our conclusions are analogous to those
for the sum of the same variables. Actually, we observed that the third and fourth raw sample 
moments slowly converge to the 
theoretical ensemble averages, i.e. for $N > 10^2$. After the crossover, the log-normal 
behavior of the MCLT is gradually recovered, as shown from the behavior of the data distributions 
as a function of $N$ in Fig.~\ref{fig:fig6} (upper panel, right plot and lower panel). We checked that the agreement between the 
data distribution and the analytical log-normal is good also along the tails for
$N = 10^3$.

To summarize, our analysis reveals that the non-Gaussian stochastic processes 
under consideration converge to the limiting distributions under both addition 
and multiplication of their jump sizes but with a rather low convergence speed, as expected by virtue of their statistical 
properties. 
All in all, our simulations mimic the actual dynamics observed in financial markets, 
where log-returns slowly converge from a leptokurtic distribution to a normal, the kurtosis 
being the parameter that controls the speed of convergence of 
the stochastic processes toward the Gaussian regime~\cite{BPBook2003,Wu2006,Figueiredo2003}.

\section{An Application to Option Pricing}
\label{op}

In the classical Black and Scholes model of option pricing~\cite{BS1973}, the SDE of Eq.~(\ref{eq:gbm}) is used 
to describe the stochastic motion of the risky underlying asset. 

Under this assumption, if we consider those options which can be exercised at the maturity only (European-style 
options), the fair value of the derivative can be expressed in an arbitrage-free market
as discounted expected value of 
the future payoff under the risk-neutral probability measure~\cite{Jondeau2006,Shreve2013,Hull1999,Glasserman2003}. Then, 
the option price reads as follows (we consider, for definitiveness, the case of call options
denoted with $C$)
\begin{equation}
C (S, t) \, = \, e^{- r \, (T - t)} \, \int_0^{+\infty} \, d S_T \, p^*_{LN} (S_T, T | S, t) \, \phi (T) \, ,
\label{eq:oc}
\end{equation}
where $S = S(t)$ is the spot price at time $t$, $S(T)$ is the asset price at the maturity $t = T$
and $\phi (T)$ generically stands for the payoff function, which will be specified in the 
following as it depends on the type of financial product. In Eq.~(\ref{eq:oc}),  
$p^*_{LN} (S_T, T | S, t)$ represents the risk-neutral log-normal conditional probability 
density, that follows from Eq.~(\ref{eq:gbm}) with the risk-free interest rate $r$ in place of 
the risky growth rate $\mu$, as a consequence of the absence of arbitrage opportunities 
and the existence of a unique risk-neutral measure in a complete market.

Note that Eq.~(\ref{eq:oc}) is particularly suited for a MC implementation~\cite{Hull1999,Glasserman2003}, as the payoff 
function is weighted by $p^*_{LN} (S_T, T | S, t)$, which is automatically recovered 
in MC simulations of Eq.~(\ref{eq:sg}) (with $r$ in place of $\mu$) since prices 
distribute according to the associated PDF.

In order to compare with the Gaussian predictions, we assume in our study that option 
prices under non-Gaussian fluctuations can be obtained as done in~\cite{Matacz2000}, i.e. 
by following the risk-neutral 
approach in exactly the same way as it is used with the GBM model. This 
strategy can be argued by using a martingale approach to option pricing
under non-Gaussian statistics~\cite{Borland2002a,Borland2002b,Borland2004}. 
Therefore, the fair value of the option contract 
can be still computed by a MC solution of Eq.~(\ref{eq:oc}), but where 
the risk-neutral dynamics of the stock prices is now driven by 
Eq.~(\ref{eq:sg}) with $\mu$ set equal to $r$ and the replacement $\xi_G \to \xi_{NG}$. 
 
 We consider in our study both plain vanilla and exotic call options, whose payoffs 
 are given by
 \begin{itemize}
 
 \item[a)] European plain vanilla call option

\begin{equation}
\phi (T) \, = \, {\rm max} \, \{ S(T) - X, 0 \} \, ,
\end{equation}
where $X$ is the strike price.
   
\item[b)] Knock-out call option
\begin{equation}
\phi (T) = \left\{
\begin{aligned}
 & \, 0 \qquad \qquad \qquad \qquad \qquad {\rm if} \, \, \exists \, \, S(t_i) \!\!: \, \, S(t_i) > U\\
 & \, {\rm max} \, \{ S(T) - X, 0 \} \qquad \, \,  {\rm elsewhere} \, ,
\end{aligned}
\right.
\end{equation}
where $U$ is a barrier price and $t_i \in [t,T]$. It is an example of path-dependent exotic option, 
since the payoff is conditional upon the underlying asset price breaching a barrier level during the option 
lifetime.
 
 \end{itemize}
 
\begin{figure}[h]
\centering
\includegraphics[width=7.5cm]{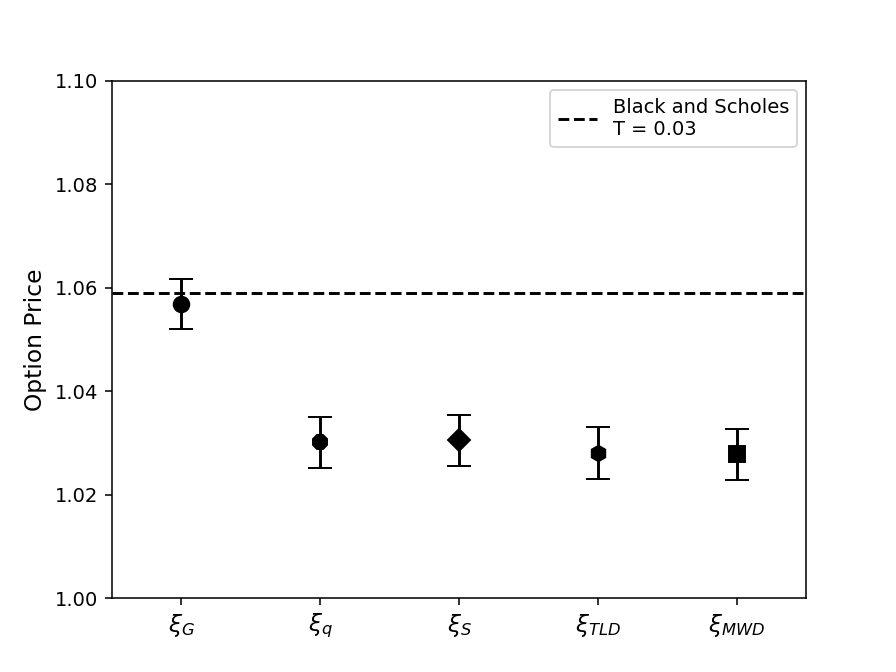}~~\includegraphics[width=7.5cm]{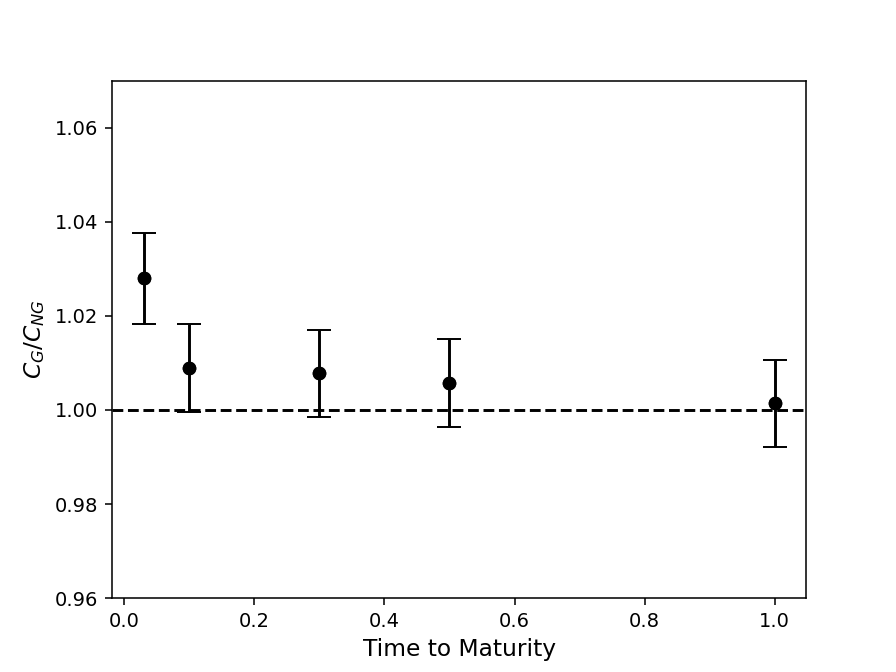}
\caption{Left panel: prices of European plain vanilla call options at-the-money for Gaussian and non-Gaussian returns, 
at short maturity $T = 0.03$. The dashed line represents the Black and Scholes 
prediction. Simulation parameters are given in the text. Right panel: the ratio of Gaussian to 
non-Gaussian option prices, as a function of the maturity, from short ($T = 0.03$) to long ($T = 1$) maturity. 
The non-Gaussian pricing is obtained by using the TLD model.}
\label{fig:fig7}
\end{figure}

\begin{figure}[h]
\centering
\includegraphics[width=7.5cm]{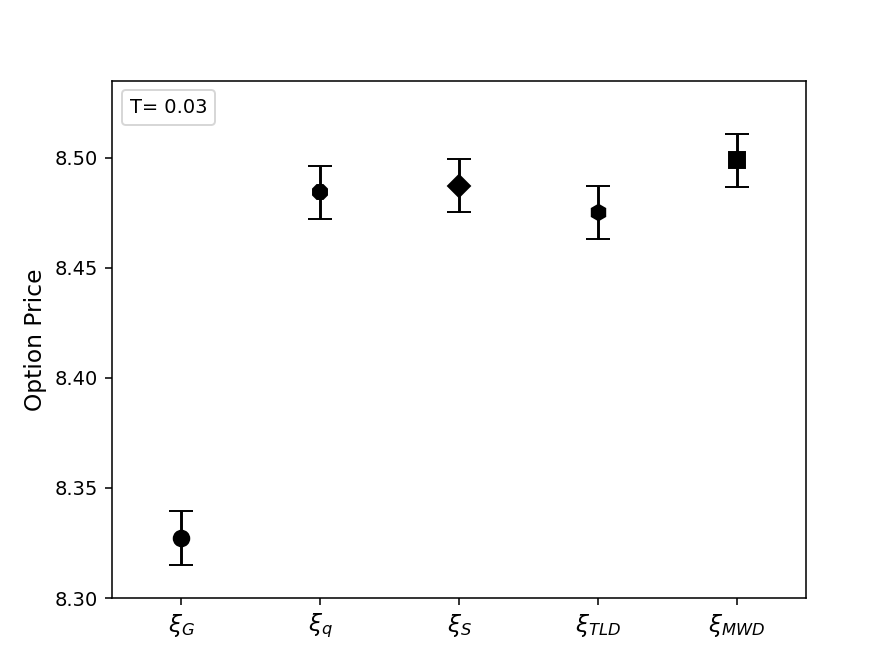}~~\includegraphics[width=7.5cm]{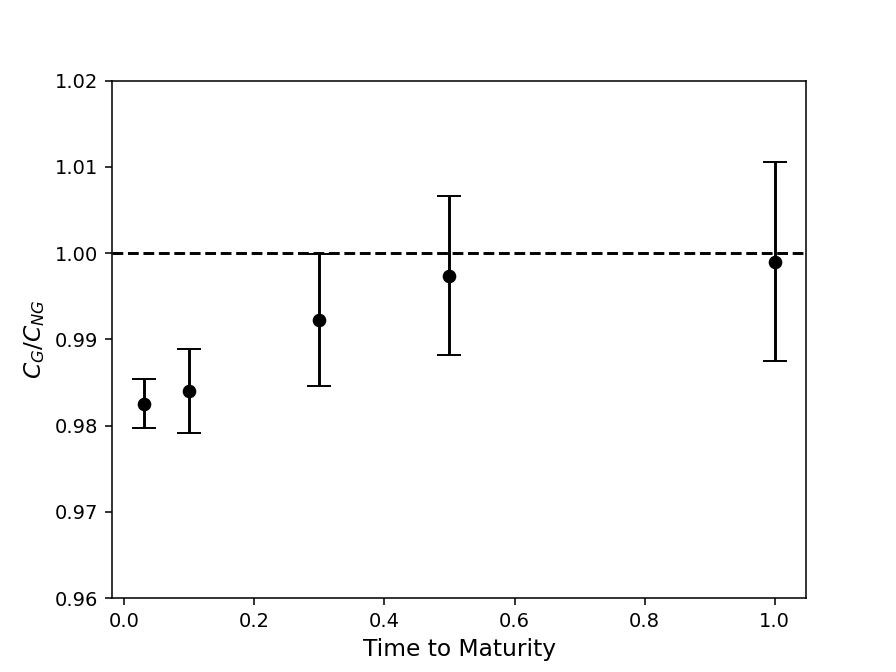}
\caption{The same as in Fig.~~\ref{fig:fig7} for knock-out call options in-the-money, with barrier price $U = 152$. 
Other simulation parameters are given in the text.}
\label{fig:fig8}
\end{figure}

A sample of the most representative results of our investigation is shown in Fig.~\ref{fig:fig7} 
(plain vanilla options) and 
Fig.~\ref{fig:fig8} (barrier options). In our simulations, we set
$\Delta t = 10^{-3}$ so that the maturity, which we vary in our study, is given by 
$T = N \Delta t$, $N$ being the number of iterations. Other parameters are:
$S(0) = 150$, $r = 0.01$ and $\sigma = 0.1$. As for the strike price, we use 
$X = 150$ in Fig.~\ref{fig:fig7} (options at-the-money) and $X = 140$ in Fig.~\ref{fig:fig8} (options in-the-money),
with barrier price $U = 152$. For all the models, we simulate $M = 10^5$ MC 
random walks and we quote the option prices with their $1\sigma$ MC error. 
 As a preliminary check of our implementation, we verified that the MC solution of 
Eq.~(\ref{eq:oc}) under Gaussian fluctuations reproduces the results of the analytical 
Black and Scholes formulae for plain vanilla options.

For both types of options, we first study the results of the 
Gaussian and non-Gaussian simulations at short maturity, in order to possibly single out 
the leptokurtic effects at small $N$. We then analyze the ratio of Gaussian to non-Gaussian 
option prices as a function of the maturity, i.e. for increasing $N$, to make contact with the 
discussion about the asymptotic convergence of the asset prices of the previous Section. 
For the latter analysis, we use the TLD modeling for non-Gaussian option pricing 
but we checked that the same pattern is present for the other distributions.

As can be seen in Fig.~\ref{fig:fig7} (left panel), the Gaussian and non-Gaussian predictions 
for short-maturity plain vanilla options display a systematic difference and the Gaussian 
approximation overprices the option at-the-money. Hence, the effect of the heavy-tailed 
distributions is to effectively reduce the Black and Scholes input volatility, in agreement with 
the behavior of the at-the-money implied volatility observed in the markets~\cite{BPBook2003}. 
At short maturity, we also noticed that the Gaussian approximation is prone to underprice the 
non-Gaussian results for out-of-money and in-the-money options, because of the contribution of fat tails.
 However, as shown in Fig.~\ref{fig:fig7} (right panel), the non-Gaussian option prices 
gradually approach the Black and Scholes results as the maturity increases. 
Actually, the longer the maturity, the better the convergence of the 
non-Gaussian asset models to the log-normal distribution as a 
consequence of the MCLT. This suggests that, according to our modeling,  
the shape of the implied volatility tends to flatten with the maturity, as 
empirically observed~\cite{BPBook2003}.

Concerning the knock-out barrier options, the results of our simulations given 
in Fig.~\ref{fig:fig8} (left panel) show that the non-Gaussian models provide 
short-maturity prices that are higher than that of the Gaussian approximation. This 
behavior has to be ascribed to the peakedness of the leptokurtic distributions 
at small $N$, that is emphasized by this kind of options where the tight pay-off 
singles out the sharp central body of the non-Gaussian returns. However, as
shown in Fig.~\ref{fig:fig8} (right panel), this effect decreases as a function of the
maturity, since the heavy-tailed distributions gradually converge to a log-normal 
and the non-Gaussian pricing tends to reproduce the same results as the standard GBM 
model.

Note, in conclusion, that the results of the different fat-tailed models 
nicely agree at the option price level, thus showing consistency also from the 
point of view of their financial applications. 

\section{Conclusions and prospects}
\label{cp}

By using large samples of synthetic data as realistic representations 
of return behavior, we performed a systematic comparative analysis 
of the most popular models 
introduced in econophysics and finance to capture the heavy-tailed, non-Gaussian 
properties of financial fluctuations in the high-frequency limit.

Our study reveals that the Student's $t$, $q$-Gaussian, truncated 
L\'evy and modified Weibull (stretched exponential) models
can explain the occurrence and scaling of large price movements observed 
in financial markets on similar footing, provided
a consistent choice of the distribution parameters is performed. As explained 
in the paper, the rationale behind the observed substantial agreement is the coherent 
use of realistic parameters as obtained in the empirical studies of high-frequency 
returns and that model the universal features of 
extreme price deviations from the Gaussian shape.

By using all the above distributions, we generated non-Gaussian random deviates 
as synthetic copies of actual financial fluctuations. We used 
these random numbers to simulate the dynamics of log-returns and asset prices
under non-Gaussian distributions and highlighted the differences with the 
standard Gaussian results. We investigated the convergence to the asymptotic
distributions of the heavy-tailed stochastic processes, to quantify their 
converge rate under addition and multiplication of i.i.d. variables.

We also presented a first application of our modeling
to option pricing, by considering both plain vanilla and path-dependent options. 
To compare with the results of the Gaussian approximation, 
we used the standard risk-neutral approach to derivative pricing and we observed differences 
between the non-Gaussian and Gaussian option prices in the limit of short maturities. 
As the maturity increases, our results gradually converge to the Gaussian 
predictions as a consequence of the limit behavior of the fat-tailed distributions.

The results obtained in the paper open the way to a number of developments and new applications.

It would be first worthwhile to test all the non-Gaussian models in comparison with
high-frequency empirical data, by using different stock market indices or traded assets, similarly 
to the comparative studies performed in \cite{BPBook2003,Eryigit2009}. 
The adherence of the models to the data could be verified by means of nonparametric one-sided tests 
such as, e.g., the Kolmogorov-Smirnov (KS) test, which tests the null hypothesis that a data sample
 is generated by a target distribution. Notably, the KS test and similar statistical tools could be used 
 to test the mutual compatibility of different models. Indeed, the two-sided variant of KS tests the null hypothesis 
 that two data samples are generated by the same distribution and it could be applied 
 on large synthetic samples such as those generated in our study.

From the point of view of modeling, a natural improvement is the inclusion
of fast diffusion observed in asset price dynamics,
 in order to study the differences
with the standard (Wiener) diffusion of GBM. Also, we are interested in taking into 
account the stochastic nature of
 volatility, which is now treated as a constant parameter. 
This would allow us to go beyond the present approximation of treating the financial 
fluctuations as independent random variables, since the inclusion of 
a stochastic volatility would drop the property of independence, as well as to develop a 
more realistic approach to derivative pricing~\cite{Fouque2000}.

Furthermore, an interesting and almost direct application of our work
would be the study of market risk measures and portfolio optimization 
under non-Gaussian return 
distributions where one can expect substantial differences 
from the Gaussian approximation because of the presence of heavy tails. 

 \section*{Acknowledgement}
 We acknowledge INFN, Sezione di Pavia, for the availability of computer resources which we
used to produce the random number samples of our study. We wish to thank 
Carlo M. Carloni Calame and Clara~L. Del Pio for helpful assistance in using such resources. 
We are also grateful to Clara L. Del Pio and Fulvio Piccinini for discussions on 
MC issues and the bootstrap method. 

\appendix
\section{Kurtosis from Non-Gaussian Monte Carlo samples}
\label{kur}

We performed the test described in this Appendix following the remarks 
raised in~\cite{DelMar2012,Celikoglu2018}.

Actually, to assess the reliability of a modeling in the tails when using non-Gaussian 
distributions, it is important comparing the 
kurtosis as computed from the generated
random samples with the theoretical results given in Section~\ref{tf}, whenever possible, as well 
as with the large but finite kurtosis typically measured in empirical studies. For the latter, 
we take as reference value the kurtosis of the S\&P 500 30 minute data quoted 
in~\cite{BPBook2003}, which is of the order of twenty. It is comparable to the kurtosis 
empirically measured for most liquid markets and most traded stocks in the intraday and daily regime, where 
it can roughly vary between ten and forty~\cite{BPBook2003,Cont2001,Jondeau2006,Platen2008,Gu2008,Tsay2010,DelMar2012}. 

As emphasized in Section~\ref{tf}, the kurtosis 
of the $\alpha = 3/2$ TLD and MWD PDFs is finite and 
given by Eq.~(\ref{eq:tldm}) and Eq.~(\ref{eq:mwdm}), respectively. On the other hand,
for our choice of the shape parameters, the theoretical kurtosis of the Student's $t$ ($\nu = 3$) and
$q$-Gaussian  ($q = 1.5$) distributions does not exist as the integral yielding the fourth moment is 
divergent. However, as remarked in Section~\ref{rd}, our non-Gaussian MC samples are generated 
using the acceptance-rejection algorithm over a wide
but bounded support [-30,+30] to simulate typically observed return fluctuations. 
Therefore, the kurtosis of our samples for the 
Student's~$t$ and $q$-Gaussian modeling is finite as well, 
like in the empirical data samples.

\begin{figure}[hbtp]
\centering
\includegraphics[width=9.5cm]{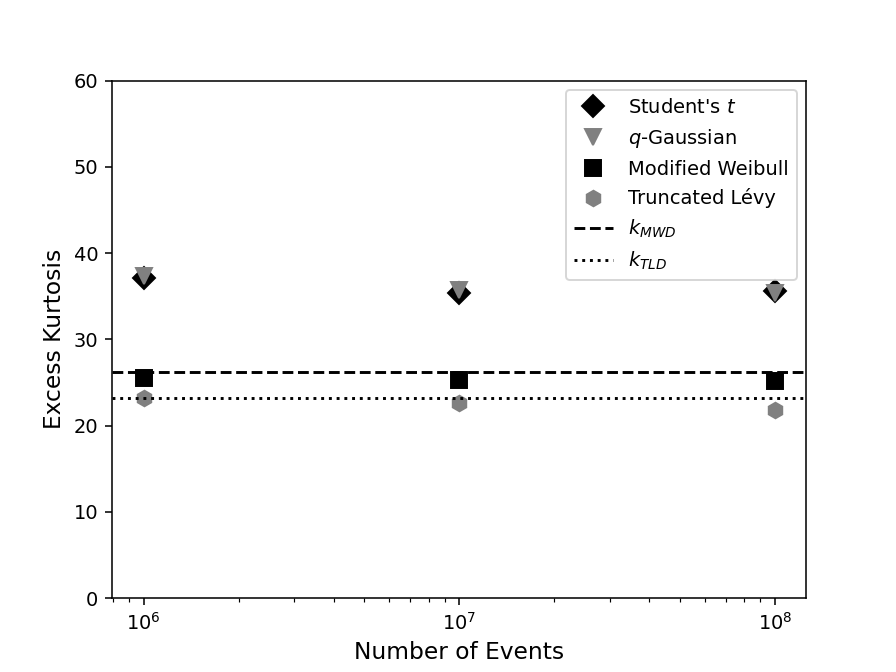}
\caption{Sample excess kurtosis values (markers) for the 
non-Gaussian return models as a function of the number of random deviates $M = 10^6,10^7,10^8$,
generated over the range [-30,+30]. The dotted and shaded lines represent the theoretical 
values of the $\alpha = 3/2$ TLD ($\lambda = 0.18$) and MWD ($c = 0.75$) modeling, respectively, 
showing agreement between the MC ensemble averages and the model predictions given by
$k_{\rm TLD} = 23.2$ and $k_{\rm MWD} = 26.2$. The sample excess
kurtosis of the Student's $t$ ($\nu = 3$) and $q$-Gaussian ($q = 1.5$) is about thirty 
five, i.e. of the same order as the empirically measured kurtosis in 
the high-frequency regime.
}
\label{fig:fig9}
\end{figure}

The results of this cross-check are shown in Fig.~\ref{fig:fig9}, where each 
excess kurtosis value is represented 
as a function of the MC sample 
size $M = 10^6, 10^7, 10^8$ generated over the range [-30,+30].
As can be seen, the sample excess kurtosis for the 
$\alpha = 3/2$ TLD ($\lambda = 0.18$) and MWD ($c = 0.75$) models 
 agrees with the 
theoretical value given by $k_{\rm TLD} = 23.2$ and $k_{\rm MWD} = 26.2$, respectively. 
On the other hand, for the equivalent return description in terms of the 
Student's $t$ and $q$-Gaussian distributions 
(with $\nu = 3$, i.e. $q = 1.5$), the sample excess kurtosis
is not surprisingly larger (being about thirty five) but of a magnitude similar to the 
empirically measured kurtosis. 
For returns in the interval [-20,+20], which is the range most 
often observed in the high-frequency limit, the sample excess kurtosis 
of the Student's $t$ and $q$-Gaussian models reduces to 
about twenty and approaches the value of the other two distributions.

For completeness, we verified that the sample kurtosis of the TLD and 
MWD distributions is independent on the sampling region, i.e. it remains stable 
and in agreement with the theoretical value, 
provided a sufficiently large interval for random number generation is used. On the 
other hand, the sample kurtosis of the Student's $t$ and $q$-Gaussian modeling 
depends on the sampling interval, as expected. Actually, we 
checked that, when using the generalized Box-M\"uller algorithm~\cite{Thistleton2007,Nelson2021} 
for sampling the $q$-Gaussian or the Student's $t$ PDFs over an unbounded 
range, the sample kurtosis of these distributions continuously grows to higher values 
as the MC sample size increases, without reaching a plateau. This behavior agrees with 
the formally infinite kurtosis for such non-Gaussian models and with the 
remarks of~\cite{Celikoglu2018}. However, in this situation, the sample kurtosis 
of the two distributions is so large that it does not match
the values observed in the markets, as previously emphasized in \cite{DelMar2012}.

\bibliography{MSNGbib}

\end{document}